\documentclass[prodmode,acmtecs]{acmsmall} 

\usepackage[ruled]{algorithm2e}

\SetAlFnt{\small}
\SetAlCapFnt{\small}
\SetAlCapNameFnt{\small}
\SetAlCapHSkip{0pt}
\IncMargin{-\parindent}

\usepackage{adjustbox}
\usepackage{array}
\usepackage{booktabs}
\usepackage{multirow}
\usepackage{rotating}

\usepackage{fancyhdr}
\newcommand*\rot{\rotatebox{90}}

\setcopyright{none}
\acmVolume{0}
\acmNumber{0}
\acmArticle{0}
\acmYear{2019}
\acmMonth{0}

\makeatletter
\def\@copyrightspace{\relax}
\makeatother

\makeatletter
\def\runningfoot{\def\@runningfoot{}}
\def\firstfoot{\def\@firstfoot{}}
\makeatother

\begin{document}

\markboth{A. Bonifati, I. Holubov\'{a}, A. Prat-P\'erez, S. Sakr}{Graph Generators: State of the Art and Open Challenges}

\title{Graph Generators: State of the Art and Open Challenges}
\author{
ANGELA BONIFATI
\affil{Lyon 1 University, France}
IRENA HOLUBOV\'{A}
\affil{Charles University, Prague, Czech Republic}
ARNAU PRAT-P\'{E}REZ
\affil{Sparsity-Technologies, Barcelona, Spain}
SHERIF SAKR
\affil{University of Tartu, Estonia}
}

\begin{abstract}
The abundance of interconnected data has fueled the design and
implementation of graph generators reproducing real-world linking
properties, or gauging the effectiveness of graph algorithms, techniques
and applications manipulating these data. We consider graph generation
across multiple subfields, such as Semantic Web, graph databases, social networks, and community detection, along with general graphs. Despite the disparate
requirements of modern graph generators throughout these communities, we
analyze them under a common umbrella, reaching out the functionalities, the
practical usage, and their supported operations.
We argue that this classification is serving the need of providing scientists, researchers and practitioners with the right data generator at hand for their work.

This survey provides a comprehensive overview of the state-of-the-art
graph generators by focusing on those that are pertinent and
suitable for several data-intensive tasks. Finally, we discuss open
challenges and missing requirements of current graph generators along with their future extensions to new emerging fields.
\end{abstract}

\maketitle

\section{Introduction}
\label{sec:intro}



Graphs are ubiquitous data structures that span a wide array of scientific disciplines
and are a subject of study in several subfields of computer science.
Nowadays, due to the dawn of numerous tools and applications manipulating
graphs, they are adopted as a rich data model in many data-intensive use
cases involving disparate domain knowledge, mathematical foundations and
computer science. Interconnected data is oftentimes used to encode domain-specific
use cases, such as recommendation networks, social networks,
protein-to-protein interactions, geolocation networks, and fraud detection analysis networks, to
name a few.

In general, we can distinguish between two broad types of graph data sets: (1) a single large graph (possibly with several components), such as social networks  or Linked Data graphs, and (2) a large set of small graphs (e.g., chemical compounds\footnote{\url{https://pubchem.ncbi.nlm.nih.gov/}} or linguistics syntax trees\footnote{\url{https://catalog.ldc.upenn.edu/ldc99t42}}). Naturally, the algorithms used in these two classes differ a lot~\cite{DBLP:books/igi/Sakr2011}. In the former case we can search, e.g., for communities and their features or shortest paths, while in the latter case we usually query for supergraphs, subgraphs, or graphs similar to a given graph pattern. In both cases, as in the other fields, quite often the respective real-world graph data is not publicly available (or simply does not exist when a particular method for data manipulation is proposed).
Even in the cases in which real data is abundant, many algorithms and techniques need to be tested on various orders of magnitudes of the graph sizes thus leading to the inception of
configurable graph generators that reproduce real-world graph properties and provide unique tuning opportunities for
algorithms and tools handling such data~\cite{sakr2013processing,sakr2016large}.

In this survey, we provide a detailed overview of the state-of-the-art
modern
graph generators by focusing on those that are pertinent and suitable for
data-intensive tasks and benchmarking context. Our aim is to cover a wide range of currently popular
areas of graph data processing. In particular, we consider graph generation
in the areas of Semantic Web, graph databases, social networks, and community detection, along with general graphs. Despite the disparate requirements of modern graph generators throughout these communities, we analyze them under a common umbrella, reaching out the functionalities, the practical usage, and the supported operations of graph generators. The reasons for this scope and classification are as follows:

\begin{enumerate}
  \item Despite the differences of the covered areas, the requirements for
modern graph data generators can be similar in particular cases. Reusing or learning from tools in other fields can thus bring new opportunities for both researchers and practitioners.
  \item The selected classification is serving the need of providing scientists, researchers, and practitioners with the right data generator at hand for their work.
\end{enumerate}

To conclude the comparative study and provide a comprehensive view of the
field,  we also overview the most popular real-world data sets used in the respective covered areas and discuss open challenges and missing requirements of modern
graph generators in view of identifying their future extensions to new emerging fields.

\paragraph*{Contributions} Our survey revisits the representatives of
modern graph data generators and summarizes their main characteristics. The detailed review
and analysis make this paper useful for stimulating new graph data generation mechanisms as well as
serving as a main technical reference for selecting suitable solutions. In particular, the introductory
categorization and comparative study enables the reader to quickly get his/her bearings in the field and
identify a subset of generators of his/her interest. Since we do not limit the survey to a particular area of graph data management,  the reader can get a broader scope in the field. Hence, while practitioners can find a solution in another previously unexpected area, researchers can identify new target areas or exploit successful results in new fields. Last but not least, 
we identify general open problems of graph data synthesis and indicate possible solutions to show that it still forms a challenging and promising research area.

\paragraph*{Differences with prior surveys}

To the best of our knowledge, this paper is the first to survey the broad
landscape of graph data generators spanning different data-intensive applications and
targeting many computer science subfields. In particular, we cover Semantic Web, graph databases, social networks, and community detection, along with general graphs. However, the literature is still lacking a comprehensive study of graph generators  for many of the specific subfields mentioned above.

A limited subset of graph database
generators, parallel and distributed graph processing generators, along
with a few of the Semantic Web data generators presented in our survey have
been discussed in a related book chapter
\cite{BFHI18} while cross-comparing them with respect to input, output,
supported workload, data model and query language, along with the
distinguished chokepoints. However, the provided classification is
inherently database-oriented.
In our work, we provide a more comprehensive and broader classification
 that serves the purpose of letting any
researcher or practitioner interested in data generation  to be able to make a better
choice of the desired graph
generator based on its functional and goal-driven features (such as the application domain,
the supported operations, and the key configuration options).
Moreover, in contrast with \cite{BFHI18}, our work encompasses graph generators of several
diverse communities, not limiting its scope to a few generators of the database
and graph processing communities.

Graph generators matching graph patterns used in data mining have been
studied in \cite{Chakrabarti:2006:GML:1132952.1132954},
focusing on mostly occurring patterns, such as power laws, size of graph diameters,
and community structure. The considered graph generators are compared in
terms of graph type, degree distributions, exponentiality, diameter, and
community effects. We refer the reader to this survey for taxonomies
involving these properties, whereas we provide here a functionality-driven
taxonomy across all the categories of graph generators that we consider.
We also point out that this survey is outdated as it does not consider the
generators that appeared in the last decade.
We fill the gap of more recent social network generators
in Section~\ref{sec:generators_socialnetworks}, as well as more recent representatives of the other categories.


Aggarwal and Subbian~\cite{AggarwalS14} have surveyed the evolution of analysis in
graphs, by primarily focusing on data mining maintenance methods and on analytical
quantification and explanation of the changes of the underlying networks.
A brief discussion on evolutionary
network data generators is carried out in
the paper. The data generation of evolutionary networks is based on the
shrinking diameters and Densification Power Law (DPL), i.e., community-guided
attachment properties~\cite{Leskovec:2005:GOT:1081870.1081893}.
Generation of graphs tackling Kronecker
recursion with recursive tensor multiplication~\cite{AkogluMF08} is then considered in the above survey.
We refer the readers to the aforementioned survey for evolutionary network generators and further
discuss the open challenges of evolving graph data in Section~\ref{sec:challenges}.


\paragraph*{Outline} The rest of the text is structured as follows: Section~\ref{sec:comparison}
provides the opening categorization and comparison of the generators. Section~\ref{sec:generators}
provides an overview of the existing graph data generators and their main features in the frame of
the proposed categories.
In Section~\ref{sec:challenges}, we highlight some of the challenges and open problems of graph
data synthesis before we conclude in Section~\ref{sec:conclusion}.

\section{Classification and Comparative Study}
\label{sec:comparison}

In order to provide a general preview of the generators and  to enable finding the target solutions easily, we start the survey with a  classification and comparative study of the existing tools. In general, there are various ways to classify them. We first offer an overview of the approaches used in state-of-the-art work after which we then introduce our approach. As mentioned in the Introduction, since this survey is unique especially in terms of scope, our classification and comparative strategy differs as well.

The graph data generators can be classified using various distinct criteria. For example,~\cite{DBLP:conf/sdm/ChakrabartiZF04} introduces two categories -- degree-based and procedural generators. In general, \emph{degree-based generators} (e.g., Barabasi-Albert model~\cite{Barabasi99emergenceScaling}) are commonly attempting to find a graph that matches it, but without providing any information about the graph or attempting to match other graph features (e.g., eigenvalues, small diameter, etc). On the other hand, \emph{procedural generators} (e.g., R-MAT~\cite{DBLP:conf/sdm/ChakrabartiZF04}) are commonly targeting simple techniques to produce graphs that are matching the characteristics of the real-world graphs (such as, e.g., the power law degree distribution).

Paper~\cite{Chakrabarti:2006:GML:1132952.1132954} introduces five categories of graph models that can be synthesized: (1) \emph{random graph models} (e.g., Erd\"{o}s-R\'{e}nyi~\cite{Erdos:1960}) generated by a random process, (2) \emph{preferential attachment models} (e.g., Barabasi-Albert model) which try to model the power law from the preferential attachment viewpoint, (3) \emph{optimization-based models} (e.g., HOT model~\cite{PhysRevLett.84.2529}) resulting from the idea that power laws can result from tolerance to risks and resource optimizations, (4) \emph{tensor-based models} (e.g., R-MAT) targeting a trade-off between low number of model parameters and efficiency, and (5) \emph{internet-specific models} corresponding to hybrids using ideas from the other categories in order to suit the specific features of the graphs.

The type of the generator can also be influenced by the benchmark involving it, whereas we can distinguish, e.g., \emph{domain-specific} benchmarks, \emph{application-specific} benchmarks, \emph{workload-driven} benchmarks,  \emph{microbenchmarks} etc.

\subsection{Classification}

At first we classify the generators on the basis of the respective application domains or user communities. In particular we distinguish (1) general graphs, (2) Semantic Web, (3) graph databases, (4) social networks, and (5) community detection. The selected classes are not rigorously defined (e.g., they are not disjoint as we will show later), but they correspond to the currently most active research areas. Thus we believe that they form a natural first acquaintance for the reader.

In Table~\ref{tab:comparisonCharacteristics}   we overview the key characteristics of the data generators clustered according to the respective application domains.\footnote{``GDBs'' stands for graph databases, ``SNs'' stands for social networks, and ``Co'' stands for community detection. Value ``$-$'' means that the information is not available or relevant.} In particular, we show:

\begin{itemize}
\item Characteristics of the \textbf{domain}:
    \begin{itemize}
      \item its \textit{type} (column ``Type''), i.e.,  fixed, specified using a schema, or extracted from input data, and
      \item the particular \textit{target} domain, or, in case of a generic tool, the chosen sample domain (column ``Target/sample'').
\end{itemize}
\item Characteristics of \textit{read}/\textit{update} \textbf{operations} (columns ``Read'' and ``Update''), i.e., whether the set of operations is fixed/generated, if it involves operation mixes (i.e., sets/sequences of operations), or if templates of operations are supported.
\item Key \textbf{configuration} options:
  \begin{itemize}
    \item whether the generator deals only with structure, or also with \emph{properties} (column ``Pro.'') of the graph (Y/N feature),
    \item supported types of \emph{distributions} (column ``Distributions'') used for generating of the data,
    \item \textit{output format} (column ``Output'') of the produced graph, and
    \item  whether the generator is \textit{distributed} (column ``Dis.'') and thus enables more efficient data generation (Y/N feature).
  \end{itemize}
\end{itemize}

\begin{sidewaystable}
\scriptsize
\centering
\tbl{Key characteristics of the generators} {
\begin{tabular}{| c | l | p{2cm} |  p{2cm} | p{1.3cm} |  l | l | p{3.2cm} | p{2cm} | l | }
 \hline
           &   & \multicolumn{2}{c}{\textbf{Domain}}
               & \multicolumn{2}{|c|}{\textbf{Operations}}
               & \multicolumn{4}{c|}{\textbf{Configuration}}
               \\ \hline
           &  \textbf{Generator}
               & \textbf{Type}
               & \textbf{Target/sample}
               & \textbf{Read}
               & \textbf{Update}
               & \textbf{Pro.}
               & \textbf{Distributions}
			   & \textbf{Output}
               & \textbf{Dis.}
               \\ \hline
\hline   
\multirow{7}{*}{\rot{\textbf{General}}}
  & Preferential attachment & -- & -- & N & N & Y & power-law & edge-list &  N  \\
\cline{2-10}
   & R-MAT & -- & -- & N & N & Y & power-law & edge-list &  N  \\
\cline{2-10}
  & HPC Scal. Graph Anal. & fixed & -- & fixed & N& N& uniform &edge-list & N   \\
\cline{2-10}
  & GraphGen & -- & -- & N & N & Y& user-defined  & node/edge-list & N   \\
\cline{2-10}
  & BTER & -- & -- &   N & N & N & user-defined & edge-list & Y  \\
\cline{2-10}
  & Darwini & -- & -- &   N & N & N & user-defined &  edge-list & Y   \\
\cline{2-10}
  & RTG & -- & -- &   N & N & N & power-law &  edge-list & N  \\
\hline
\hline 
\multirow{19}{*}{\rot{\textbf{Semantic Web}}}
 & LUBM & fixed & university  & fixed & N & Y & random (LCG) &  RDF & N   \\
\cline{2-10}
 & LBBM & extracted & Lehigh university BibTeX  & N & N & Y & Monte Carlo &  RDF & N   \\
\cline{2-10}
 & UOBM & fixed & university  & fixed & N & Y & random &  RDF & N   \\
\cline{2-10}
 & IIMB & fixed & movies  & N & N & Y & random &  RDF & N   \\
\cline{2-10}
 & BSBM & fixed & e-commerce  & fixed & N & Y & mostly normal &  RDF, relational & N   \\
\cline{2-10}
 & SP$^2$Bench & fixed & DBLP  & fixed & N & Y & based on DBLP  & RDF & N   \\
\cline{2-10}
 & \cite{Duan:2011:AOC:1989323.1989340} & extracted & -- & N & N & Y & -- &  RDF & N    \\
\cline{2-10}
 & DBPSB & extracted & DBpedia &  templates & N & Y & random &  RDF & N   \\
\cline{2-10}
 & LODIB & fixed & e-commerce &  N & N & Y & 44 types &  RDF & N   \\
\cline{2-10}
 & Geographica & fixed & OpenStreetMap  & fixed + templates  & N & Y & -- &  RDF & N   \\
\cline{2-10}
 & WatDiv & schema-driven & user-defined  & templates & N & Y & uniform, normal, Zipfian &  RDF & N   \\
\cline{2-10}
 & RBench & extracted & DBLP, Yago  & templates & N & Y & from real-world data &  RDF & N  \\
\cline{2-10}
 & S2Gen & schema-driven & social network & Y & N & N & user-defined & RDF & N     \\
\cline{2-10}
 & RSPLab & schema-driven & agnostic & Y & Y & N & user-defined & RDF & N     \\
\cline{2-10}
 & LDBC SPB & fixed & media  & mixes & N & Y & power-law, skewed values, value correlation &  RDF & N  \\
\cline{2-10}
 & LinkGen & schema-driven & user-defined & templates & N  & N & Gaussian, Zipfian & RDF & N\\
\hline
\hline  
\multirow{4}{*}{\rot{\textbf{GDBs}}}
  & XGDBench & fixed  & social network  & generated & generated & Y & power-law &  MAG &  Y  \\
\cline{2-10}
  & gMark & schema-driven &  user-defined  & generated &  N  & Y & uniform, normal, Zipfian &  N-triples & N    \\
\cline{2-10}
  & graphGen & pattern-driven & user-defined  & -- & -- & Y & -- &  GraphJson, CypherQueries & N   \\
\hline
\hline 
\multirow{8}{*}{\rot{\textbf{SNs}}}
 & \cite{Barrett:2009:GAL:1995456.1995598} & fixed & social network & N & N & Y & simulation-driven & impl. NA &  -- \\
\cline{2-10}
 & \cite{Yao2011} & fixed & social network & N & N & N & power-law & impl. NA & --  \\
\cline{2-10}
 & LinkBench & fixed & social network & generated & generated & Y & Facebook & impl. NA & -- \\
\cline{2-10}
 & S3G2 & fixed & social network  & N & N  & Y & Facebook  & CSV, RDF & Y   \\
\cline{2-10}
 & SIB & fixed & social network &  mixes & mix & Y & from real-world data &  RDF & N   \\
\cline{2-10}
 & \cite{Sukthankar-SocialInfo2014} & schema-driven & social-network & N & N & Y & power-law & CSV & N   \\
\cline{2-10}
 & LDBC SNB  & fixed & social network &  generated & generated  & Y & Facebook &  CSV, RDF & Y     \\
\cline{2-10}
  & \cite{Nettleton2016} & schema-driven & social network & N & N & Y & power-law & impl. NA & --  \\
\hline
\hline   
\multirow{4}{*}{\rot{\textbf{Co}}}
  & \cite{danon2005comparing} & -- & -- & Y & N & N & uniform & edge-list &  N    \\
\cline{2-10}
  & LFR & -- & -- & Y & N & N & power-law & edge-list  & N      \\
\cline{2-10}
  & LFR-Overlapping & -- & -- & Y & N & N & power-law & edge-list & N     \\
\cline{2-10}
  & Stochastic Block Models & -- & -- & Y & N & N & user-defined & edge-list & Y     \\
\hline
\end{tabular} }
\label{tab:comparisonCharacteristics}
\end{sidewaystable}

\paragraph{Number of Generators, Size and Output Format of the Data}
As we can see, the biggest set of available generators can be found in the Semantic Web application domain, probably due to its recent popularity and a number of research groups dealing with this topic. None of these generators is natively implemented in a distributed manner and thus primarily generating  output at the Big Data scale. On the other hand, the \emph{Linked Open Data} (LOD) is expected to be large in general; however, this is the case of the whole \emph{LOD Cloud}\footnote{\url{https://lod-cloud.net/}}, but not necessarily of the particular data sets forming it.

The second large group of generators also corresponds to a popular application domain -- social networks. In this case the size of the output graph is important if we require realistic features of the result. However, surprisingly many of the proposals do not provide an implementation at all and amongst the others there is a high percentage of those that are not highly scalable.

In case of the other application-specific domains the amount of generators is relatively small. As we will show in Section~\ref{sec:overlapping}, the situation is not so critical. Some of the application-specific generators can be re-used also in other application domains or the general graph generators can be used.

Considering the output format of the data, as expected, the generators produce data in a standard format (e.g., RDF) or in a reasonable graph-related form (e.g., edge list).

\paragraph{Domain}

If we consider the features of the particular domain of the generators, in most cases it is expectably \emph{fixed}, i.e., it is pre-defined (describing, e.g., a social network) and cannot be modified. It is the simplest option, but it does mean that the generator is simple too -- it can still focus on other complex aspects of the output. While the \emph{schema-driven} approaches enable to influence the target domain using a user-supplied schema, there also exist approaches where the domain is \emph{extracted} from sample data. The particular target (in case of fixed) or sample (in case of schema-driven or extracted) domains do not provide very rich areas. They either correspond to the respective application domains (like in the case of social networks), or they are based on well-known and commonly used data sets (such as, e.g., DBLP~\cite{dblp} or DBpedia~\cite{Bizer:2009:DCP:1640541.1640848}). Except for general graphs without any domain, in all other cases we can find a flexible representative with schema-driven/pattern-driven domain or a domain extracted from sample data.

\paragraph{Operations}
In the case of operations, most commonly the respective generators are accompanied with a set of fixed read operations or sequences of operations (query mixes) representing typical behavior of a user. In some cases this aspect is more flexible -- either query templates are used or the operations are generated, e.g., in order to access most of the data in the generated graph. On the other hand, update operations are provided only in a small amount of cases. As we will discuss in Section~\ref{sec:evolving}, the more general problem of evolving graphs is still an open issue.

\paragraph{Configuration}
A natural feature of the generators is to provide as realistic graphs as possible. Hence, most of them focus not only on the structure of the output graph, but also the properties. For the purpose of generating graphs with near real-world characteristics various distributions are used, such as power-law, Zipfian etc. Especially interesting are distributions extracted from real-wold data (such as, e.g., Facebook in case of the social network domain).

\subsection{Overlapping}
\label{sec:overlapping}

As we have mentioned, the basic classification of the generators that we have used in this paper is relatively vaguely based on the current application domains or research areas. In addition, some of the generators are either general, and thus can be used universally, or have features applicable in more than one domain/research area. So the classes we use can overlap, as depicted in Table~\ref{tab:overlapping}.

\begin{table}
\scriptsize
\centering
\tbl{Overlapping of classes of generators} {
\begin{tabular}{| c | l | l | l | l | l | l | l | }
 \hline
           &  \textbf{Generator}
               & \textbf{\rot{General}}
               & \textbf{\rot{Semantic Web}}
               & \textbf{\rot{Graph databases\ }}
               & \textbf{\rot{Social networks}}
               & \textbf{\rot{Community detection\ }}
               \\ \hline
\hline   
\multirow{6}{*}{\rot{\textbf{General}}}
  & Preferential attachment & {\bf x} & & &  &  \\
\cline{2-7}
   & R-MAT    & {\bf x} & & &  &  \\
\cline{2-7}
   & HPC Scal. Graph Anal.  & {\bf x} & & &  &  \\
\cline{2-7}
  & GraphGen  & {\bf x} & & x &  & \\
\cline{2-7}
  & BTER      & {\bf x} & & &  & \\
\cline{2-7}
  & Darwini   & {\bf x} & & &  & \\
\cline{2-7}
  & RTG   & {\bf x} & & &  & x \\
\hline
\hline 
\multirow{16}{*}{\rot{\textbf{Semantic web}}}
 & LUBM  & & {\bf x} & &  &  \\
\cline{2-7}
 & LBBM  & & {\bf x} & &  &  \\
\cline{2-7}
 & UOBM  & & {\bf x} & &  &  \\
\cline{2-7}
 & IIMB & & {\bf x} & &  & \\
\cline{2-7}
 & BSBM & & {\bf x} & &  &  \\
\cline{2-7}
 & SP$^2$Bench & & {\bf x} & &  &  \\
\cline{2-7}
 & \cite{Duan:2011:AOC:1989323.1989340} & & {\bf x} & &  &  \\
\cline{2-7}
 & DBPSB & & {\bf x} & &  &  \\
\cline{2-7}
 & LODIB & & {\bf x} & &  &  \\
\cline{2-7}
 & Geographica & & {\bf x} & &  &  \\
\cline{2-7}
 & WatDiv & & {\bf x} & &  &  \\
\cline{2-7}
 & RBench & & {\bf x} & &  &  \\
\cline{2-7}
  & S2Gen & & {\bf x} & &  &  \\
\cline{2-7}
  & RSPLab & & {\bf x} & &  &  \\
\cline{2-7}
 & LDBC SPB & & {\bf x} & &  &  \\
\cline{2-7}
 & LinkGen & & {\bf x} & &  &  \\
\hline
\hline  
\multirow{3}{*}{\rot{\textbf{GDBs}}}
  & XGDBench & & & {\bf x} & x  &  \\
\cline{2-7}
  & gMark & & x & {\bf x} & x  &  \\
\cline{2-7}
  & graphGen & & & {\bf x} &  &  \\
\hline
\hline 
\multirow{8}{*}{\rot{\textbf{SNs}}}
 & \cite{Barrett:2009:GAL:1995456.1995598} & & & & {\bf x}  & \\
\cline{2-7}
 & \cite{Yao2011}  & & & & {\bf x}  & \\
\cline{2-7}
 & LinkBench  & & & x & {\bf x}  & \\
\cline{2-7}
 & S3G2  & & x & x & {\bf x}  & \\
\cline{2-7}
 & SIB & & x & & {\bf x}  &  \\
\cline{2-7}
 & \cite{Sukthankar-SocialInfo2014}  & & & & {\bf x}  & \\
\cline{2-7}
 & LDBC SNB   & & x & x & {\bf x}  & \\
\cline{2-7}
  & \cite{Nettleton2016}  & & & & {\bf x}  &  \\
\hline
\hline   
\multirow{4}{*}{\rot{\textbf{Co}}}
  & \cite{danon2005comparing} & x & & & &  {\bf x} \\
\cline{2-7}
  & LFR & x & & & &  {\bf x}  \\
\cline{2-7}
  & LFR-Overlapping & x & & & &  {\bf x} \\
\cline{2-7}
  & Stochastic Block Models & x & & & &  {\bf x} \\
\hline
\end{tabular} }
\label{tab:overlapping}
\end{table}

For example, many general or domain-agnostic graph generators, such as the
preferential attachment~\cite{Barabasi99emergenceScaling} or R-MAT, are typically used to test graph analytics frameworks when large real graphs are not available. 

Similarly, some social network graph generators such as LDBC SNB, S3G2 or
LinkBench, can be used to test graph databases. In the case of the first two,
even though they are designed not to be specific to any type of technology,
the graph databases are their main target.  Additionally, they also provide serializers for
RDF, thus they can also be used to test RDF systems.

In the case of LinkBench, nothing prevents the user to load the generated graph
in a (graph) database (e.g., Facebook uses MySQL in~\cite{Armstrong:2013:LDB:2463676.2465296}) to
test a workload similar to Facebook and extend and complement it with more
graph queries like those in LDBC SNB.

Generators for community detection aim, in general, at creating graphs with a
more realistic structure (graphs with communities of nodes where the density of
edges is larger internally than externally). Even though these generators are,
in general, used to test community detection algorithms (they generate also the
expected communities in the graph), some studies also use them for general graph
purposes or to test graph analytics algorithms besides community detection.

\section{Graph Data Generators}
\label{sec:generators}

In this section, we discuss the various graph data generators based on the classification introduced before in more detail. For each category, we first describe the key features of each of the representative examples and summarize their strengths and weaknesses. The goal is to offer a detailed information about each of the tools in the context of its competitors from the same domain.

\subsection{General Graphs}
\label{sec:generators_general}

We start by focusing on approaches that have been designed for dealing with the
generation of general graph data that is not aimed at a particular application
domain. In general, such generators focus on reproducing properties observed in
real graphs regardless of their domain such as the degree distribution, the
diameter, the presence of a large connected component, a community structure or
a significantly large clustering coefficient.


\paragraph {Preferential Attachment} Barabasi and
Albert~\cite{Barabasi99emergenceScaling} introduced a graph generation model
that relies on two main mechanisms. The first mechanism is continuously
expanding the graphs by adding new vertices. The second mechanism is to
preferentially attach the new vertices   to the nodes/regions that are already
well connected. So, in this approach, the generation of large graphs is governed
by standard, robust self-organizing mechanisms that go beyond the
characteristics of individual applications.

\paragraph {R-MAT} R-MAT (\emph{R}ecursive \emph{Mat}rix) is a procedural
synthetic graph generator which is designed to generate power-law degree
distributions~\cite{DBLP:conf/sdm/ChakrabartiZF04}.  The generator is recursive
and employs a fairly small number of parameters.  In principle, the strategy of
this generator is to achieve simple means to produce graphs whose properties correspond to
properties of the real-world graphs. In particular, the design goal of R-MAT is to
produce graphs which mimics the degree distributions, imitate a community
structure and have a small diameter. R-MAT can generate weighted, directed and
bipartite graphs.

\paragraph{HPC Scalable Graph Analysis Benchmark} The HPC Scalable Graph
Analysis Benchmark~\cite{HPCgraph,Bader:2005:DIH:2099301.2099360} consists of a
weighted, directed graph that has a power-law distribution and four related
analysis techniques (namely graph construction, graph extraction with BFS,
                     classification of large vertex sets, and graph analysis
                     with betweenness centrality). The generator has the
following parameters: the number of nodes, the number of edges, and maximum weight of
an edge. It outputs a list of tuples containing identifiers of vertices of an
edge (with the direction from the first one to the second one) and weights
(positive integers  with a uniform random distribution) assigned to the edges of
the multigraph.  The algorithm of the generator is based on R-MAT.

\paragraph{GraphGen} For the purpose of testing the scalability of an indexing
technique called FG-index~\cite{Cheng:2007:FTV:1247480.1247574} on the size of the
database of graphs, their average size and average density, the authors have
also implemented a synthetic generator called
GraphGen\footnote{https://www.cse.ust.hk/graphgen/}. It relies on data generation code for associations and sequential
patterns provided by IBM\footnote{From 1996, no longer available at
\url{http://www.almaden.ibm.com/cs/projects/iis/hdb/Projects/data
mining/mining.shtml}}. GraphGen yields a collection of undirected, labeled and
connected graphs. It addresses the performance evaluation of frequent subgraph mining
algorithms and graph query processing algorithms. The result is represented as a
list of graphs, each consisting of a list of nodes along with a list of edges.


\paragraph{BTER} BTER (Block Two-Level
Erd\"{o}s-R\'{e}ny)~\cite{kolda2014scalable} is a graph generator based on the
creation of multiple Erd\"{o}s-R\'{e}ny graphs with different connection
probabilities  of which they are connected randomly between them. As the main feature, BTER is able
to reproduce input degree distributions and average clustering
coefficient per degree values. The generator starts by grouping the vertices
by degree $d$, and forming groups of size $d+1$ of nodes with degree $d$. Then, these
groups are assigned an internal edge probability in order to match the observed
average clustering coefficient of the nodes of such degree. Based on this
probability, for each node, the excess degree (i.e, the degree that in
expectation will not be realized internally in the group) is computed and used to connect
nodes from different groups at random.
The authors report that BTER is able to generate graphs with billions of edges.

\paragraph{Darwini} Darwini~\cite{edunov2016darwini} is an extension of BTER
designed to run on Vertex Centric computing frameworks like
Pregel~\cite{malewicz2010pregel} or Apache Giraph~\cite{ching2015one}, with the
additional feature that it is more accurate when reproducing the clustering
coefficient of the input graph. Instead of just focusing on the average
clustering coefficient for each degree, Darwini is able to model the clustering
coefficient distribution per degree. It achieves this by gathering the nodes
of the graph into buckets based on the expected number of closed triangles that they need
to close in order to attain the expected clustering coefficient. The latter is
sampled from the input distributions. Then, the vertices in each bucket are
connected randomly with a probability that would produce the expected
desired number of triangles for each bucket. Then, as in BTER, the excess degree
is used to connect the different buckets. The authors report that Darwini is
able to generate graphs with billions and even trillions of edges.


\paragraph{RTG} \emph{The Random Typing Generator} (RTG)~\cite{DBLP:journals/datamine/AkogluF09} aims at generating realistic graphs. In particular, it outputs (un)weighted, (un)directed, as well as uni/bipartite graphs, whereas the realism of the output is ensured by 11 laws (e.g., densification power law, weight power law, small and shrinking diameter, community structure, etc.) known to be typically exhibited by real-world graphs. On input it requires 4 parameters ($k$, $q$, $W$, and $\beta$) that correspond to the core Miller's observation~\cite{miller1957} that a random process (namely, having a keyboard with $k$ characters and a space, the process of random typing of $W$ words, where the probability of hitting a space is $q$ and the probability of hitting any other characters is $(1-q)/k$) leads to Zipf-like power laws (of the resulting words) plus in addition (using imbalance factor $\beta$) ensure homophily and community structure. RTG is based on the idea creating an edge between pairs of consecutive words.

Paper~\cite{DBLP:conf/wbdb/AmmarO13} further extends this idea mainly in the direction of simplification of specifying of the parameters. Instead of Miller's parameters that are not much associated with graphs, the authors prove and exploit their relationship with size and density of the target graph.

\paragraph{Strengths and Weaknesses of General Graph Generators}
In general, existing general graph generators produce graphs with the following
properties: skewed degree distribution (e.g., power law), small diameter, a large largest connected component, large
clustering coefficient, and some degree of
community structure. Degree distribution can be typically configured while other
properties are just a result of the generation process and cannot be controlled
by any means. This is not the case in the work of BTER and Darwini, which besides the degree distribution,
they also allow tuning of the clustering coefficient. However, some use cases
demand for more control on the characteristics of the generated graphs.
This is the case, for example, of benchmarking, where the underlying graph
structure has direct implications to the performance of the graph algorithms
to run. For this reason, the design of graph generators with the capability of fine
tuning the characteristics of the generated graphs still remains as an open
challenge. The questions that remain are what characteristics to tune
and which are the algorithms that depend on such characteristics.

\subsection{Semantic Web}
\label{sec:generators_LinkedData}
With the dawn of the concept of Linked Data it is a natural development that there would emerge respective
benchmarks involving both synthetic data and  data sets   with real-world characteristics.
The used data sets correspond to RDF representation of relational-like data~\cite{Guo2005158,Bizer09theberlin}, social network-like data~\cite{Schmidt2010}, or specific and significantly more complex data structures such as biological data~\cite{Wu2014}. In this section, we provide an overview of benchmarking systems involving a kind of graph-based RDF data generator or data modifier. 

\paragraph{LUBM} The use-case driven Lehigh University Benchmark (LUBM)\footnote{\url{http://swat.cse.lehigh.edu/projects/lubm/}} considers the university domain. The ontology defines 43 classes and 32 properties~\cite{Guo2005158}. In addition, 14 test queries are provided in the LUBM benchmark. In particular, the benchmark focuses on \emph{extensional} queries, i.e., queries which target the particular data instances of the ontology, as an opposite to \emph{intentional} queries, i.e., queries which target properties and classes of the ontology. The Univ-Bench Artificial  (UBA) data generator features repeatable and random  data generation (exploiting classical linear congruential generator, LCG, of numbers). In particular, the data which is produced by the generator are assigned zero-based indexes (i.e., \emph{University0}, \emph{University1} etc.), thus they are reproducible at any time with the same indexes.  The generator naturally allows to specify a seed for random number generation, along with the starting index and the desired number of universities.

An extension of LUBM, the Lehigh BibTeX Benchmark (LBBM)~\cite{Wang2005}, enables generating synthetic data for different ontologies. The data generation process is managed through two main phases: (1) the property-discovery phase, and (2) the data generation phase. LBBM provides a probabilistic model that can emulate the discovered properties of the data of a particular domain and generate synthetic data exhibiting similar properties. Synthetic data are generated using a Monte Carlo algorithm. The approach is demonstrated on the Lehigh University BibTeX ontology which consists of 28 classes along with 80 properties. The LUBM benchmark includes 12 test queries that were designed for the benchmark data. Another extension of LUBM, the University Ontology Benchmark (UOBM)\footnote{\url{https://www.cs.ox.ac.uk/isg/tools/UOBMGenerator/}}, focuses on two aspects: (1) usage of all constructs of OWL Lite and OWL DL~\cite{owl} and (2) lack of necessary links between the generated data which thus form isolated graphs~\cite{Ma:2006:TCO:2094613.2094629}. In the former case the original ontology is replaced by the two types of extended versions from which the user can choose. In the latter case cross-university and cross-department links are added to create a more complex graph.

\paragraph{IIMB} Ferrara et al.~\cite{Ferrara08OM} proposed the ISLab Instance Matching Benchmark (IIMB)\footnote{\url{http://www.ics.forth.gr/isl/BenchmarksTutorial/}} for the problem of instance matching. For any two objects $o_1$ and $o_2$ adhering to different ontologies or to the same ontology, instance matching is specified in the form of a function $Om(o_1, o_2) \rightarrow \{0, 1\}$,  where $o_1$ and $o_2$ are linked to the same real-world object (in which case the function maps to $1$) or $o_1$ and $o_2$ are representing different objects (in which case the function maps to $0$). It targets the domain of movie data which contains 15 named classes, along with 5 objects and 13 datatypes. The data are extracted from IMDb\footnote{\url{http://www.imdb.com/}}. The data generator corresponds to a data modifier which simulates differences between the data. In particular it involves data value differences (such as typographical errors or usage of different standard formats, e.g., for names), structural heterogeneity (represented by different levels of depth for properties, diverse aggregation criteria for properties, or missing values specification) and logical heterogeneity (such as, e.g., instantiation on disjoint classes or various subclasses of the same superclass).

\paragraph{BSBM} The Berlin SPARQL Benchmark (BSBM)\footnote{\url{http://wifo5-03.informatik.uni-mannheim.de/bizer/berlinsparqlbenchmark/}}, is centered around an e-commerce application domain with object types such as \emph{Customer}, \emph{Vendor}, \emph{Product} and \emph{Offer} in addition to the relationship among them~\cite{Bizer09theberlin}.
The benchmark provides a workload that has 12 queries with 2 types of query workloads (i.e., 2 sequences of the 12 queries) emulating the navigation pattern and search of a consumer seeking a product. The data generator is capable of producing arbitrarily scalable datasets by controlling the number of products ($n$) as a scale factor.  The scale factor also impacts other data characteristics, such as, e.g., the depth of type hierarchy of products, branching factor, the number of product features,  etc. BSBM can output two representations, i.e. an RDF representation along with a relational representation. Thus, BSBM also defines an SQL~\cite{sql} representation of the queries. This allows comparison of SPARQL~\cite{sparql} results  to be compared against the performance of traditional RDBMSs.

\paragraph{SP$^2$Bench} The SP$^2$Bench\footnote{\url{http://dbis.informatik.uni-freiburg.de/forschung/projekte/SP2B/}}
is a language-specific benchmark~\cite{Schmidt2010} which is based on the DBLP dataset. 
The generated datasets follow the key characteristics of the original DBLP dataset. In particular, the data mimics the correlations between entities. All random functions of the generator use a fixed seed that ensures that the data generation process is deterministic. SP$^2$Bench is accompanied by 12 queries covering the various types of operators such as RDF access paths in addition to typical RDF constructs.


\paragraph{DBPSB} DBpedia SPARQL Benchmark (DBPSB)\footnote{\url{http://aksw.org/Projects/DBPSB.html}} proposed at the University of Leipzig has been designed using workloads that have been generated by applications and humans~\cite{Morsey2011,Morsey:2012:UBR:2900929.2901031}. In addition, the authors argue that benchmarks like LUBM, BSBM, or SP$^2$Bench resemble relational database benchmarks involving relational-like data which is structured using a small amount of homogeneous classes, whereas, in reality, RDF datasets are tending to be more heterogeneous. For example, DBpedia 3.6 consists of 289,016 classes, whereas 275 of them are defined based on the DBpedia ontology. In addition, in property values different data types as well as references to objects of the various types are
used. Hence, they presented a universal SPARQL benchmark generation approach which uses a flexible data production mechanism that mimics the input data source. This dataset generation process begins using an input dataset; then multiple datasets with different sizes  are generated by duplicating all the RDF triples with changing their namespaces.  For generating smaller datasets, an adequate selection of all triples is selected randomly or using a sampling mechanism over the various classes in the dataset. The methodology is applied on the DBpedia SPARQL endpoint and a set of 25 templates of SPARQL queries is derived to cover frequent SPARQL features.

\paragraph{LODIB} The Linked Open Data Integration Benchmark (LODIB)\footnote{\url{http://lodib.wbsg.de/}} has been designed with the aim of reflecting the real-world heterogeneities that exist on the Web of Data in order to enable testing of Linked Data translation systems~\cite{DBLP:conf/www/RiveroSBR12}. It provides a catalogue of 15 data translation patterns (e.g., rename class, remove language tag etc.), each of which is a common data translation problem in the context of Linked Data. The benchmark provides a data generator that produces three different synthetic data sets that need to be translated
by the system under test into a single target vocabulary. They  reflect the pattern distribution in analyzed 84 data translation examples from the LOD Cloud. The data sets reflect the same e-commerce scenario used for BSBM.

\paragraph{Geographica} The Geographica benchmark\footnote{\url{http://geographica.di.uoa.gr/}} has been designed to target the area of geospatial data~\cite{DBLP:conf/semweb/GarbisKK13} and respective SPARQL extensions GeoSPARQL~\cite{battle2012enabling} and stSPARQL~\cite{koubarakis2010modeling}. The benchmark involves a real-world workload that uses openly available datasets that cover various geometry elements (such as, e.g., lines, points, polygons, etc.) and  a synthetic workload. In the former case there is a (1) a micro benchmark that evaluates primitive spatial functions (involving 29 queries) and (2) macro benchmark that tests the performance of RDF engines in various application scenarios such as  map exploring and search (consisting of 11 queries). In the latter case of a synthetic workload the generator produces synthetic datasets of different sizes that corresponds to an ontology based on OpenStreetMap  and instantiates query templates. The generated SPARQL query workload is corresponding to spatial joins and selection using 2 query templates.

\paragraph{WatDiv} The Waterloo SPARQL Diversity Test Suite (WatDiv)\footnote{\url{http://dsg.uwaterloo.ca/watdiv/}} has
been designed at the University of Waterloo. It implements stress testing tools that focus on addressing the observation
that the state-of-the-art SPARQL benchmarks do not fully cover the variety of queries and
workloads~\cite{Aluc:2014:DST:2717213.2717229}. The benchmark focuses on two types of query aspects --
structural and data-driven -- and performs a detailed analysis on existing SPARQL benchmarks
(LUBM, BSBM, DBPSB, and SP$^2$Bench) using these two properties of queries. The structural features involve triple
pattern count,  join vertex degree, and join vertex count. The data-driven features involve
result cardinality and several types of selectivity. The analysis of the four benchmarks reveals that their diversity is
insufficient for evaluation of the weaknesses/strengths of the distinct design alternatives implemented by the different
RDF systems.

In particular, WatDiv, provides (1) a data generator which generates scalable datasets according to the WatDiv
schema, (2) a query template generator which produces a  set of query templates according to the WatDiv schema, and
(3) a query generator that uses the generated templates and instantiates them  with real RDF values from the dataset, and (4) a feature extractor which extracts the structural features of the generated data and workload. 

\paragraph{RBench} RBench~\cite{Qiao:2015:RAR:2723372.2746479} is an application-specific benchmark which receives any RDF dataset as an input and produces a set of datasets, that have similar characteristics of the input dataset, using size scaling factor $s$ and (node) degree scaling factor $d$. These factors ensure that the original RDF graph $G$ and the synthetic graph $G'$ are similar and the average node degree and the number of edges of $G'$ are changed by $s$ and $d$ respectively. A query generation process has been implemented to produce 5 different types of queries (edge-based queries, node-based queries, path queries, star queries, subgraph queries) for any generated data. The benchmark project FEASIBLE~\cite{Saleem2015} is also an application-specific benchmark; however, contrary to RBench, it is designed to produce benchmarks from the set of sample input queries of a user-defined
size.

In practice, one way for handling big RDF graphs is to process them using the
\emph{streaming} mode where the data stream could consist of the edges of the
graph. In this mode, the RDF processing algorithms can process the input
stream in the order it arrives while using only a limited amount of
memory~\cite{mcgregor2014graph}. The streaming mode has mainly  attracted the attention of the
RDF and Semantic Web community.

\paragraph{S2Gen}   Phuoc et al.~\cite{le2012linked} presented
an evaluation framework for linked stream data processing engines. The framework
uses a dataset generated with the Stream Social network data Generator
(S2Gen), which
simulates streams of user interactions (or events) in social networks
(e.g., posts) in addition to the  user metadata such as users' profile
information, social network relationships, posts, photos and GPS information.
The data generator of this framework provides the users the flexibility to
control the characteristics of the generated stream by tuning a range of
parameters, which includes the frequency at which interactions are generated,
limits such as the maximum number of messages per user
and week, and the correlation probabilities between the different objects (e.g.,
users) in the social network.

\paragraph{RSPLab} Tommasini et al.~\cite{tommasini2017rsplab} introduced
another framework for benchmarking RDF Stream Processing systems, RSPLab. The
Streamer component of this framework is designed to publish RDF streams from the
various existing RDF benchmarks (e.g., BSBM, LUBM).
In particular, the Streamer  component uses TripleWave\footnote{\url{http://streamreasoning.github.io/TripleWave/}}, an
open-source framework which enables to share RDF streams on the
Web~\cite{mauri2016triplewave}.   TripleWave acts as a means for plugging-in
and combining streams from multiple Web data sources using either pull or push mode.

\paragraph{LDBC}  The Linked Data Benchmark Council\footnote{\url{http://ldbcouncil.org/industry/organization/origins}} (LDBC)~\cite{Angles:2014:LDB:2627692.2627697} 
had the goal of developing open source, yet industrial grade benchmarks for RDF and graph databases.
In the Semantic Web
domain, it released the Semantic Publishing Benchmark (SPB)~\cite{spb} that has been inspired by the
Media/Publishing industry (namely BBC\footnote{\url{http://www.bbc.com/}}). The application scenario
of this benchmark simulates a media or a publishing organization that handles large amount
of streaming content (e.g., news, articles). The data generator mimics three types of relations in the generated synthetic data:
correlations of entities, data clustering, and random tagging of entities. Two workloads are provided: (1) basic, involving an interactive query-mix querying the relations between entities in reference data, and (2) advanced,  focusing on interactive and analytical query-mixes. The LDBC has designed two other benchmarks: the Social Network Benchmark (SNB)~\cite{Erling:2015:LSN:2723372.2742786} for the social network domain  (see Section~\ref{sec:generators_socialnetworks}) and Graphalytics~\cite{Iosup:2016:LGB:3007263.3007270}   for the analytics domain.

\paragraph{LinkGen} LinkGen is a synthetic linked data generator that has been designed to generate RDF datasets for a given vocabulary~\cite{10.1007/978-3-319-46547-0_12}. The generator is designed to receive a vocabulary as an input  and supports two statistical distributions for generating entities:  Zipf's power-law distribution and Gaussian distribution. LinkGen can augment the generated data with inconsistent and noisy  data such as updating a given datatype property with two conflicting values or  adding triples with syntactic errors. The generator also provides a feature to inter-link the generated objects with real-world ones from user-provided real-world datasets. The datasets can be generated in any of of two modes: on-disk and streaming.

\paragraph{Strengths and Weaknesses of Semantic Web Graph Generators}  Graphs are intuitive and standard representation for the RDF model that form the basis for the Semantic Web community which has been very active on building several benchmarks, associated with graph generators that had various design principles.

A comparison of 4 RDF benchmarks (namely TPC-H~\cite{TPC-H} data expressed in RDF, LUBM, BSBM, and SP$^2$Bench) and 6 real-wold data sets (such as, e.g.,  DBpedia, the Barton Libraries Dataset~\cite{barton-benchmark} or
WordNet~\cite{Miller:1995:WLD:219717.219748}) has been reported by~\cite{Duan:2011:AOC:1989323.1989340}. The authors focus mainly on the  \emph{structuredness} (\emph{coherence}) of each benchmark dataset claiming that a primitive metric (e.g., the number of triples or the average in/outdegree) quantifies only some target characteristics of each dataset. With respect to a type $T$ the degree of structuredness of a dataset $D$  is based on  the regularity of instance data in $D$ in conforming to type $T$. The type system is extracted from the data set by finding the RDF triples that have property\footnote{\url{http://www.w3.org/1999/02/22-rdf-syntax-ns\#type}}   and extract type $T$ from their object. Properties of $T$ are determined as the union of all properties of type $T$. The structuredness is then expressed as a weighted sum of share of set properties of each type, whereas higher weights are assigned to types with more instances. The authors show that the structuredness of the chosen benchmarks is fixed, whereas real-world RDF datasets are belonging to the non-tested area of the spectrum. As a consequence, they introduce a new benchmark that receives as input any dataset associated with a required level of structuredness and size (smaller than the size of the original data), and exploits the input documents as a seed to produce a subset of the original data with the target structuredness and size. In addition, they show that structuredness and size mutually influence each other.

With the recent increasing momentum of streaming data, the Semantic Web community started to consider the issues and challenges of RDF streaming data. However, there are still a lot of open challenges that need to tackled in this direction such as covering different real-world application scenarios.

\subsection{Graph Databases}
\label{sec:generators_GraphDatabases}

Currently there exists a number of papers which compare the efficiency of graph databases with regards to distinct use cases, such as  the community detection problem~\cite{Beis2015}, social tagging systems~\cite{Giatsoglou2011}, graph traversal~\cite{Ciglan:2012:BTO:2374486.2375242}, graph pattern matching~\cite{Pobiedina2014}, data provenance~\cite{Vicknair:2010:CGD:1900008.1900067}, or even several distinct use cases~\cite{Grossniklaus2013Towar-24253}. However, the number of graph data generators and benchmarks that have been designed specifically for graph data management systems (Graph DBMS)  is relatively small. Either a general graph generator is used for benchmarking graph databases, such as, e.g., the HPC Scalable Graph Analysis Benchmark~\cite{Dominguez-Sal:2010:SGD:1927585.1927590} or the graph DBMS benchmarking tools are designed while having in mind a more general scope. Hence it is questionable whether a benchmark  that is targeted specifically for graph databases is necessary. \cite{Dominguez-Sal:2010:DDG:1946050.1946053} discussed this question and related topics. On the basis of a review of applications of graph databases (namely, social network analysis,  genetic interactions and recommendation systems), the authors analyzed and discussed the features of the graphs for these types of applications and how such features can affect the benchmarking process, various types of operations used in these applications and the characteristics of the evaluation setup of the benchmark. In this section, we focus on graph data generators and benchmarks that have been primarily targeting graph DBMSs.

\paragraph{XGDBench} XGDBench~\cite{Dayarathna:2014:GDB:2676904.2676939} is an extensible  benchmarking platform for graph databases used in cloud-based systems. Its intent is to automate
benchmarking of graph databases in the cloud by focusing on the domain social networking services. It extends the Yahoo! Cloud Multiplicative Attribute (MAG) Graph Serving Benchmark (YCSB)~\cite{Cooper:2010:BCS:1807128.1807152} and provides a set of standard workloads representing various performance issues. In particular, the workload of XGDBench involves basic operations such as read / insert / update / delete an attribute, loading of the list of neighbours and BFS traversal. Using the generators, 7 workloads are created, such as update heavy, read mostly, short range scan, traverse heavy etc.
The data model of XGDBench is a simplified version of the Multiplicative Attribute Graph (MAG)~\cite{Kim2010} model, a synthetic graph model which models the interactions between node attributes and  graph structure.
The generated graphs are thus in MAG format, with power-law degree distribution closely simulating real-world social networks.
The simplified MAG algorithm accepts the required number of nodes and for each node the number of attributes, a threshold for random  initialization of attributes, a threshold for edge affinity which determines the existence of an edge between two nodes, and an affinity matrix. 
Large graphs can be generated on multi-core systems by XGDBench multi-threaded version.

\paragraph{gMark}  gMark~\cite{gMark} is a schema-driven and domain-agnostic generator of both graph instances and graph query workloads. It can generate instances under the form of N-triples and queries in various concrete query languages, including OpenCypher\footnote{\url{https://neo4j.com/developer/cypher-query-language/}}, recursive SQL, SPARQL and LogicQL. In gMark, it is possible to specify a \emph{graph configuration} involving the admitted edge predicates and node labels occurring in the graph instance along with additional parameters such as degree distribution, occurrence constraints, etc. The \emph{Query workload configuration} describes parameters of the query workload to be generated, by including the number of queries, arity, shape and selectivity of the queries.
The problem of deciding whether there exists a graph that satisfies a defined graph specification $G$ is NP-complete. The same applies to the problem of deciding
whether there exists a query workload compliant with a given query workload configuration $Q$. In view of this, gMark adopts a best effort approach in which the
parameters specified in the configuration files are attained in a relaxed fashion in order to achieve linear running time whenever possible.


\paragraph{GraphGen}  GraphAware GraphGen\footnote{\url{http://graphgen.graphaware.com/}} is a graph generation engine based on Neo4j's\footnote{\url{https://neo4j.com/}} query language OpenCypher~\cite{GraphGen}.  It creates nodes and relationships based on a schema definition expressed in Cypher, and it can also generate property values on both
nodes and edges. As such, GraphGen is a precursor of property graphs generators. The resulting graph can be exported to several formats (namely GraphJson\footnote{\url{https://github.com/GraphAlchemist/GraphJSON/wiki/GraphJSON}} and CypherQueries) or loaded directly to a DBMS. However, it is very likely that it is not maintained anymore due to the lack of available recent commits.

\paragraph{Strengths and Weaknesses of Graph Database Generators}
The graph DBMS generators discussed in this section have in common the fact that they can generate semantically rich labeled graphs with properties (ranging from properties values in GraphGen to MAG structures in XGDBench). They are also capable of generating graph instances and query workloads in
concrete syntaxes (among which OpenCypher in GraphGen and gMark) and one of them (XGDBench) can also handle update operations on both graph structure and content. However, more comprehensive graph DBMS generators that also produce data manipulation operations (such as updates for graph databases) are urgently needed. Additionally, none of these generators is enabled to work on corresponding query languages for property graphs, such as the newly emerging standard GQL~\cite{gql-2018} and G-Core~\cite{AnglesABBFGLPPS18}. Hence, a full-fledged graph DBMS generator for property graphs and property graph query workloads~\cite{BFVY18} is still missing and there exists an interesting opportunity to build such a generator in the near future.

Another apparent inconvenience is represented by the fact that explicit correlations among graph elements cannot be encoded for instance in gMark or GraphGen, whereas they could be fruitful in order to reproduce the behavior of real-world graphs in which attribute values are correlated one with another. On the other hand, social network and Linked Data generators that support correlations (as highlighted in Section \ref{sec:generators_socialnetworks} and Section \ref{sec:generators_LinkedData}) typically exhibit a fixed schema and are not not necessarily multi-domain as are some of the graph DBMS generators discussed in this section (namely GraphGen and gMark).

\subsection{Social Networks}
\label{sec:generators_socialnetworks}

On-line social networks, like Facebook, Twitter, or LinkedIn, have become a
phenomenon used by billions of people every day and thus providing extremely
useful information for various domains. However, an analysis of such type of
graphs has to cope with two problems: (1) availability of the data and (2)
privacy of the data. Hence, data generators which provide realistic synthetic
social network graphs are in a great demand.

In general, any analysis of social networks identifies their various specific
features~\cite{Chakrabarti:2006:GML:1132952.1132954}. For example, a social
networks graph often has a high  degree of
transitivity of the graph (so-called \emph{clustering coefficient}). Or, its diameter, i.e., the longest shortest path
amongst some fraction (e.g. 90\%) of all connected nodes, is usually low due to
weak ties joining faraway cliques.

Another key aspect of social networks is the community effect. A detailed
study of structure of communities in 70 real-world networks is provided, e.g.,
in~\cite{Leskovec:2008:SPC:1367497.1367591}.
~\cite{Prat-Perez:2014:CSS:2621934.2621942} analyzed the structure of
communities (clustering coefficient, triangle participation ratio,
diameter, bridges, conductance, and size) in both real-world graphs and outputs of
existing graph generators such as LFR~\cite{PhysRevE.78.046110} and the
LDBC SNB~\cite{Erling:2015:LSN:2723372.2742786}. They found out that discovered
communities  in different graphs have common distributions and that communities
of a single graph have different characteristics and are challenging to be represented using a single
model.

The existing social network generators try to reproduce different aspects of the
generated network. They can be categorized into statistical and agent-based.
\emph{Statistical
approaches}~\cite{PhysRevE.78.046110,Yao2011,Armstrong:2013:LDB:2463676.2465296,Pham2013,Sukthankar-SocialInfo2014,Erling:2015:LSN:2723372.2742786,Nettleton2016}
focused on reproducing aspects of the network. In \emph{agent-based
approaches}~\cite{Barrett:2009:GAL:1995456.1995598,Bernstein:2013:SAS:2499604.2499609}
the networks are constructed by directly simulating the agents' social choices.


\paragraph{Realistic Social Network}
~\cite{Barrett:2009:GAL:1995456.1995598} focused on the construction of
realistic social networks. For this purpose the authors combine both
private and public data sets with large-scale agent-based techniques. The process works as
follows: In the first step it generates  synthetic data by combining public and
commercial databases. In the second step, it determines a set of activity
templates. A 24-hour activity sequence including geolocations is assigned to
each synthetic individual. To demonstrate the approach, the authors create a
synthetic US population consisting of people and households together with
respective geolocations. For this purpose the authors combine simulation and
data fusion techniques utilizing various real-world data sources such as U.S.
Census data, responses to a time-use survey or an activity survey.
The result is captured by a dynamic network of social contacts. Similar methods for
agent-based strategies have been reported
in~\cite{Bernstein:2013:SAS:2499604.2499609}.

\paragraph{Linkage vs. Activity Graphs}
\cite{Yao2011} distinguished between two types of social network graphs -- the
\emph{linkage graph}, where nodes correspond to people and edges correspond to their
friendships, and the \emph{activity graph}, where nodes also represent people
but edges correspond to their interactions. On the basis of the analysis of
Flickr\footnote{\url{https://www.flickr.com/}} social links and
Epinions\footnote{\url{http://www.epinions.com/}} network of user interactions,
the authors discover that they both exhibit high clustering coefficient
(community structure), power-law degree distribution and small diameter.
Considering the dynamic properties they both have relatively stable clustering
coefficient over time and follow the densification law. On the other hand,
diameter shrinking is not observed in Epinions activity graph and there is a
difference in degree correlation (i.e., frequency of mutual connections of
similar nodes) -- activity graphs have neutral, whereas linkage graphs have positive degree correlation. With regards to the findings, the
proposed generator focuses on linkage graphs with positive degree correlation.
For this purpose it extends the forest fire spreading process
algorithm~\cite{Leskovec:2005:GOT:1081870.1081893} with link symmetry. It has
two parameters: the \emph{symmetry
probability} $P_s$ and the \emph{burning probability} $P_b$. $P_b$ ensures a forward burning process based on BFS in which
fire burns strongly with $P_b$ approaching 1.  $P_s$ ensures backward linking
from old nodes to new nodes and ``adds fuel to the fire as it brings more
links''. 

\paragraph{LinkBench} The LinkBench
benchmark~\cite{Armstrong:2013:LDB:2463676.2465296} has been designed for the
purpose of analysis of efficiency of a database storing Facebook's production
data. The benchmark considers true Big Data and related problems with sharding,
replication etc. The social graph at Facebook comprises objects (nodes with IDs,
version, timestamp and data) and associations (directed edges, pairs of node
IDs, with visibility, timestamp and data). The size of the target graph is the
number of nodes. Graph edges and nodes are generated concurrently during bulk
loading. The space of node IDs is divided into chunks which enable parallel
processing. The edges of the graph are generated in accordance with the results
of analysing  real-world Facebook data (such as outdegree distribution). A
workload corresponding to 10 graph operations (such as insert object, count the
number of associations etc.) and their respective characteristics over the
real-world data is generated for the synthetic data.

\paragraph{S3G2} The Scalable Structure-correlated Social Graph Generator
(S3G2)~\cite{Pham2013} is a general framework which produces a directed labeled
graph whose vertices represent objects having property values. The respective
classes determine the structure of the properties. S3G2 does not aim at
generating near real-world data, but at generating synthetic graphs with a
correlated structure. Hence, the existing data values influence the probability
of choosing a particular property value from a pre-defined dictionary or connecting two nodes. For example, the degree distribution can be
correlated with the properties of a node and thus, e.g., people who have many
friend relationships typically post more comments and pictures. The data
generation process starts with generating a number of nodes with property values
generated according to specified property value correlations and then adding
respective edges according to specified correlation dimensions. It has multiple
phases, each focusing on one correlation dimension. Data is generated in a Map
phase corresponding to a pass along one correlation dimension. Then the data are
sorted along the correlation dimension in the following Reduce phase. A
heuristic observation that ``the probability that two nodes are connected is
typically skewed with respect to some similarity between the nodes'' enables to
focus only on sliding window of most probable candidates. The core idea of the
framework is demonstrated using an example of a social network (consisting of
persons and social activities).  The dictionaries for property values are
inspired by DBpedia and provided with 20 property value correlations. The edges
are generated according to 3 correlation dimensions.

\paragraph{SIB} The developers of the Social Network Intelligence BenchMark (SIB)\footnote{\url{https://www.w3.org/wiki/Social_Network_Intelligence_BenchMark}} based the design of their benchmark on the claim that the state-of-the-art benchmarks are limited in reflecting the characteristics of the real RDF databases and are mostly focusing on the relational style aspects. Hence, they proposed a benchmark for  query evaluation using real graphs~\cite{sib}. The proposed benchmark mimics using an RDF store for a social network. The distribution of the generated data for each type follows the  distribution of the associated type inferred from real-world social networks. Additionally, association rules are exploited for representing the real-world data correlation in the generated synthetic data. The  generated data is linked with the RDF datasets from DBpedia. The benchmark specification contains 3 query mixes -- interactive, update, and analysis -- expressed in SPARQL 1.1 Working Draft.

\paragraph{Cloning of Social Networks} ~\cite{Sukthankar-SocialInfo2014}
introduces two synthetic generators to reproduce two characteristics typically
observed in social networks: node features and multiple link types. Both
generators extend  the generator proposed by~\cite{wang2011leveraging}.
which starts with a small number of
nodes and new nodes are added until the network reaches the required number. It
has two basic parameters: homophily and link density. A high \emph{homophily}
value reflects that links have higher chances to be established among the nodes belonging to the same community,
whereas the community membership is represented by the same labels.

The first proposed generator is Attribute Synthetic Generator (ASG), used for
reproducing the node feature distribution of standard networks and rewiring the
network to preferentially connect nodes that exhibit a high feature similarity.
The network is initialized with a group of three nodes. New nodes and links
are added to the network based on link density, homophily, and feature
similarity. As new nodes are created, their labels are assigned based on the
prior label distribution. After the network has reached the same number of nodes
as the original social media dataset, each node initially receives a random
attribute assignment. Then a stochastic optimization process is used to move the
initial assignments closer to the target distribution extracted from social
media dataset using the Particle Swarm Optimization algorithm. The tuned
attributes are then used to add additional links to the network based on the
feature similarity parameter -- a source node is selected randomly and connected
to the most similar node. The second proposed generator, so-called Multi-Link Generator
(MLG), further  uses link co-occurrence statistics from the original dataset to
create a multiplex network. MLG uses the same network growth process as ASG.
Based on the link density parameter, either a new node is generated with a label
based on the label distribution of the target dataset or a new link is created
between two existing nodes.

\paragraph{LDBC SNB} Despite having a common Facebook-like dataset, thanks to
three distinct workloads the Social Network Benchmark
(SNB)~\cite{Erling:2015:LSN:2723372.2742786} provided by LDBC represents three
distinct benchmarks. The network nodes correspond to people and the edges
represent their friendship and messages they post in discussion trees on their
forums. The three query workloads involve: (1) SNB-Interactive, i.e., complex
read-only queries accessing a high portion of data, (2) SNB-BI, i.e.,
queries accessing a high percentage of  entities and grouping them in various
dimensions, and (3) SNB-Algorithms, involving graph analysis algorithms, such as
community detection, PageRank, BFS, and clustering. The graph generator, called
Datagen, is a fork of \texttt{S3G2} ~\cite{Pham2013} and realizes power laws, uses skewed
value distributions, and ensures reasonable correlations between graph
structures and property values. Additionally, it extends \texttt{S3G2} with "spiky"
patterns in the distribution of social network activity along the timeline, also
provides the ability of generating update streams to the social network. Datagen
is also based on Hadoop in order to provide scalability, but compared to \texttt{S3G2}, it
contains numerous performance improvements and the ability to be deterministic
regardless of the number of computer nodes used for the generation of the graphs
and for a given set of configuration parameters.


\paragraph{Towards More Realistic Data} \cite{Nettleton2016} argued that the majority of existing works focuses on
topology generation which approximates the features of a real-world social
network (e.g., community structures, skew degree
distribution, a small average path length, or a small graph diameter); however,  this is usually done without any data. Hence, they
introduced a general stochastic modeling approach that enables the users to
fill a graph topology with data. The approach has three steps: (1) topology
generation (using R-MAT) plus community identification using the Louvain
method~\cite{1742-5468-2008-10-P10008} or usage of a real-world topology from
SNAP\footnote{\url{https://snap.stanford.edu/data/}}, (2) data definition
that describes  definitions of attribute values (distribution profiles) using a
parameterizable set of affinities and data propagation rules, and (3) data
population.

\paragraph{Strengths and Weaknesses of Social Network Generators}
Compared to more general graph generators, social network generators focus
mainly on reproducing intra- and inter-node feature correlations.
Among existing generators, LDBC SNB and S3G2 look
like the most advanced ones in terms of the complexity of the generated graph
and the amount of features and correlations they can generate, while providing a
large degree of scalability. Their generation process is based on input dictionaries and have configuration files that
allow tweaking  many parameters of the generated graphs, but their schema is mainly
static and cannot be easily configured to meet the needs of other use cases
besides the benchmarks they have been designed for. In this regard, the approaches
like those proposed in~\cite{Nettleton2016} and~\cite{Sukthankar-SocialInfo2014} offer a
more flexible and understandable configuration process to tweak the types,
values, and correlations between different features.

Regarding the correlation between the underlying graph structure and the node
features, approaches such as LDBC SNB, S3G2 or ~\cite{Nettleton2016}
take into account this aspect and the generated graphs have realistic
structural properties while similar nodes have a larger probability of being
connected. However, their approach seems to be more based on
intuition and common sense than to be backed up by  any study of how the
relation between structure and attributes showcase in real social networks. In
this regard, this remains as a clear open challenge for social network
generators.

Finally, scalability is another aspect to be considered in social network graph
generators. LDBC SNB and S3G2 are engineered with this in mind, thus they
provide a way to scale to billions of nodes and edges. This is not the case for
the other generators, which can make them impractical if our goal is to generate
real sized social network graphs.

\subsection{Testing Community Detection}
\label{sec:generators_community_detection}

Community detection is one of the many graph analytics algorithms typically used
in domains such as social networks or bioinformatics. \emph{Communities} are usually defined as sets of nodes that are highly mutually connected, while being scarcely connected to
the other nodes of the graph. Such communities emerge from the fact that real-world graphs
are not random, but follow real-world dynamics that make similar entities to
have a larger probability to be connected. As a consequence, detected
communities are used to reveal vertices with similar characteristics, for
instance to discover functionally equivalent proteins in protein-to-protein
interaction networks, or persons with similar interests in social networks. Such
applications have made community detection a hot topic during the last 15
years with tens of developed algorithms and detection
strategies~\cite{doi:10.1002/wics.1403,Kim:2015:CDM:2854006.2854013}. For
comparing the quality of the different proposed techniques, one needs graphs
with \emph{reference communities}, that is, communities known beforehand. Since
it is very difficult to have large real-world graphs with reference communities
(mainly because these would require a manual labeling), graphs for benchmarking
community detection algorithms are typically generated synthetically.

\paragraph{Danon et al.} The first attempts to compare community detection algorithms using synthetic
graphs proposed the use of random graphs composed by several Erd\"{o}s-R\'{e}nyi
subgraphs, connected more internally than externally~\cite{danon2005comparing}.
Each of these subgraphs has the same size and the same internal/external density
of edges. However, such graphs miss the realism observed in real-world graphs, where
communities are of different sizes and densities, thus several proposals exist
to overcome such an issue.

\paragraph{LFR} Lancichinetti, Fortunato
and Radicchi (hence LFR)~\cite{PhysRevE.78.046110} propose a class of benchmark
graphs for community detection where communities are of diverse sizes and
densities. The generated communities follow a power-law distribution whose parameters can be configured. The degree of the
nodes is also sampled from a power-law distribution. Additionally, the generator
introduces the concept of the ``mixing factor'', which consists of the percentage of
edges in the graph connecting nodes that belong to distinct communities. Such parameter
allows  the degree of modularity of the generated graph  to be tuned, thus
testing the robustness of the algorithms under different conditions. The
generation process is implemented as an optimization process starting with an empty graph and
progressively filling it with nodes and edges guided by the specified constraints.

\paragraph{LFR-Overlapping} Lancichinetti, Fortunato and
Radicchi~\cite{PhysRevE.80.016118} extended LFR to support the notion of
directed graphs and overlapping communities. Overlapping communities extend the
notion of communities by allowing the sharing of vertices, thus a vertex can
belong to more than one community. This extended generator allows controlling
the same parameters of LFR, as well as the amount of overlap of
the generated communities.

\paragraph{Stochastic Block Models} Another popular family of generators
widely used in the community detection field are the stochastic block
models~\cite{holland1983stochastic}. In such models, the community structure of
the graph is typically defined as an array of $n$ community or cluster sizes
and a density square matrix of size $n \times n$ containing the density of
intra-cluster edges (in the diagonal of the matrix) and the density of
inter-cluster edges. Then, a stochastic procedure is run to sample graphs
from such array and matrix, using the sizes to compute the possible edges
and the densities as probabilities of such edges to exist. The popularity of
these methods stem from its simplicity and scalability, which makes them
suitable for generating large graphs fast and in distributed environments,
provided that the density matrix is sparse (as it happens in most of real-world graphs). Moreover, given
that the generation process of such models is mathematically tractable, they
are typically used to analyze the limitations of algorithms for community detection
 such as those based on modularity
optimization~\cite{fortunato2007resolution} or based on
triads~\cite{prat2016put}. Extensions of such models exist, such as the
Mixed Membership Stochastic Block Model~\cite{airoldi2008mixed}, for
overlapping communities.

\paragraph{Strengths and Weaknesses of Community Detection Generators}
Besides synthetic graph generators, Yang and Leskovec~\cite{yang2015defining}
proposed the use of real-world graphs with explicit group annotations (e.g.,
forums in a social network, categories of products, etc.) to infer what they
call \emph{meta-communities}, and use them to evaluate overlapping community
detection algorithms. However, a recent study from Hric, Darst and
Fortunato~\cite{hric2014community} reveal a loose correspondence between
communities (the authors refer to them as \emph{structural communities}) and
meta-communities.  This result reveals that  algorithms working for structural
communities do not work well for finding meta-communities and vice versa,
suggesting significantly different underlying characteristics
between the two types of communities, which are yet to be
identified.

In this regard and to the best of our knowledge, there are no available
generators that can generate graphs with meta-communities for community detection
algorithm benchmarking. The closest one is the LDBC SNB data
generator which has been provided by the
generation of groups of users in the social network. Even though the generation
process does not specifically enforce the generation of groups
(meta-communities) for benchmarking community detection algorithms, the study~\cite{Prat-Perez:2014:CSS:2621934.2621942} reveals that these groups are more similar
to the real meta-communities than those structural communities generated by the
LFR benchmark.

The differences observed between structural and meta-communities reveal the need
of more accurate community definitions that are more tight and more specific to the domain or the
use case. Current community detection algorithms and graph generators for
community detection are stuck to the traditional (and vague) definition of
community, assuming that there exists a single algorithm that would fit all the
use cases. Thus, future work requires the study of domain-specific community
characteristics that can be used to generate graphs with a community structure
that accurately resembles that of specific use cases, and thus revealing which
are the best algorithms for each particular scenario.

\section{Challenges and Open Problems}
\label{sec:challenges}

To conclude the  overview of the state-of-the-art of graph data generation, in this section we discuss several of the open challenges.

\subsection{Simple Usage, Simple Parameters}
The proposal of a data generator (not necessarily for graph data) has to face an
important schism. On one hand, it must provide the user with as
many parameters as possible in order to enable him/her to generate
arbitrary data.
This approach seems to be reasonable, but it entails a shortcoming due to
the fact that ordinary users are
unwilling to use complex benchmarking tools. This observation can be
seen, for example, in the case of XML benchmarks -- even though there exist
robust and complex data generators (such as
ToXGene~\cite{conf/webdb/BarbosaMKL02}, which supports the specification of
structural aspects, value distributions, references etc.), the most popular
benchmarking tool is XMark~\cite{Schmidt:2002:XBX:1287369.1287455}, which models
a single use case and enables its users to specify just the size of the data. Hence, the
other extreme is to provide a simple data generator which does not require any
complex settings and thus guarantees a simple and fast benchmarking process.

Considering the complex structure of graph data and the variety of applications
requiring highly specific types of graphs, the latter solution is difficult to
implement. A reasonable compromise can be found in a data generator which is
provided with sample graph data and is capable of automatic analysis of its
structural and value features in order to learn the complex parameters. We could see this type approach in some cases, such as Semantic Web generators LBBM or DBPSB.

\subsection{Large Scale Graphs with Realistic Structure}

Most of existing graph generators are focused on generating large graphs with realistic
structural characteristics and focus principally on reproducing the degree
distribution and the clustering
coefficient~\cite{kolda2014scalable,edunov2016darwini}. However, there are other
structural characteristics that one might be interested in reproducing for a large
graph, such as the diameter, the size of the largest connected component, or the
hierarchical community structure. Graph practitioners are highly interested in knowing
how other high-level structural characteristics affect the performance of graph
queries and graph algorithms. Hence, a compelling open challenge consists
of creating
graph generators that allow one to reproduce diverse structural characteristics
of the graphs along with large scale sizes.

\subsection{Single- vs. Multi-Domain}

Most of existing graph generators also generate graphs that are either not labeled or
are specific to a given domain (e.g.,  social networks). Graphs from different
domains have different schemas, structural characteristics, property
distributions, etc. which might have an impact on the performance of the
application under test. Thus, graph processing engine developers are asking for
generators or tools to generate multi-domain graphs  in a flexible and holistic manner, allowing to configure aspects
such as size, schemata, data distributions and other structural
characteristics such as degree distributions, clustering coefficients, and
so on.

\subsection{Generating Noisy Graphs and Graphs with Anomalies}
Injecting noise and/or anomalies and errors into graphs is crucial for
testing both machine learning algorithms working on this complex data and
data quality techniques aiming at detecting anomalies and repairing graph
data.

Concerning the former, analyzing and labeling structural networks is
deemed to be more difficult for graph datasets in the presence of noise.
Since de-noising graph data is difficult to achieve, various machine learning-based
approaches have been adapted to work with noise (i.e., mislabeled
samples) or outliers, such as
imbalanced graph classification \cite{PanZ13} and
binary graph classification with positive and negative weights \cite{CheungSML16}.
Synthetic graph generators that take into account noisy and missing data
have been studied in \cite{NamataG10}, where graph identification is presented in
order to  model the inference of a cleaned output network from a
noisy input graph.
Concerning the latter, data quality techniques handling graph data are recently considering ad-hoc
generation of graph data and graph quality rules in order to evaluate the
effectiveness of error detection and data repairing algorithms \cite{FanWX16a,AriouaB18}. The
corresponding graph quality rules are typically handcrafted by domain
experts, whereas an automatic generation of such rules along with the graph
data generation in tandem would be an interesting future challenge for the
community.

\subsection{Streaming Graph Generators}
Stream computing is a new paradigm that is necessitated by various modern data generation scenarios such as the ubiquity of mobile devices, location services,
sensor pervasiveness and emerging IoT applications. These applications generate the data with high Velocity, one of the
main 3V characteristics of Big Data applications~\cite{sakr2016big}. In most of these high speed data generation scenarios,
various objects are connected together with different relations and data exchanges  in a graph-structured manner. The
Semantic Web community has been considering the aspect of implementing streaming RDF generator and benchmarks; however, there is still a clear lack on considering this aspect in other important and timely domains such as IoT. 
In addition, graph streaming generators should  consider some specific aspects for the stream processing domains such as the out-of-order handling (late arrivals)~\cite{li2008out} and the variety in the schemas and formats of the different data streaming sources. It is also recommended for the streaming graph generators to support the distributed environment as this is the most common scenario for such type of applications.

\subsection{Evolving Graph Data}
\label{sec:evolving}
As user requirements as well as environments change, most of the existing
applications naturally evolve over time, at least to some extent. This evolution
usually influences the structure of the data and consequently all the related
parts of the application (i.e., storage strategies, operations, indexes etc.).
In the world of graph data such graphs that change with time are denoted as \emph{evolving}, \emph{temporal}, \emph{dynamic}, or \emph{time-varying}. They can be modeled as labeled graphs, where the labels capture some measure of time~\cite{Michail2015}.

The evolution of graphs can be considered from multiple perspectives. We can
assume a static set of nodes and a varying set of edges.  Or, there are
applications where the graphs only ``grow'', i.e., the set of nodes and/or edges
is only extended with new items. In the most general case we can assume any
changes in both set of nodes and set of edges. Anyway with the evolution aspect
the complexity of classical graph problems increases
significantly~\cite{Michail2015,Wu:2014:PPT:2732939.2732945}. In some graph
applications, such as, e.g., social networks, the evolution of the data is a
significant aspect, especially in the activity
graphs~\cite{Kumar:2006:SEO:1150402.1150476,doreian1997evolution,Hellmann2014583,wang2013,Viswanath:2009:EUI:1592665.1592675,Kossinets88}.
However, as shown
in~\cite{Leskovec:2005:RMT:2101235.2101254,Leskovec:2005:GOT:1081870.1081893},
evolving graphs have further specific features. For example, some graphs grow
over time according to a \emph{densification power law} which means that in real
graphs, edges tend to appear at a higher pace than vertices, meaning that these
graphs densify as they grow. Also the way the new edges are distributed has the
effect of a shrinking diameter that ends up stabilizing as the graph grows with time.

A related problem is \emph{data versioning} and its respective ability to query across multiple versions of data or to carry out general analysis.
This problem has been investigated for instance within the domain of Linked
Open Data~\cite{DBLP:conf/semweb/Papakonstantinou16,DBLP:conf/esws/MeimarisP16,fernandez2015towards,fernandez2015bear}.

The respective data generator should hence be able to simulate a natural growth
and/or changes in the structure of the graph with regards to the various
features of distinct use cases. However,  even though the area of dynamic graphs
is intensively studied, surprisingly there seem to exist only very few proposals
of a generator for dealing with this area.
In~\cite{GoerkeKlugeSchumm2012_1000029825} the authors focus on \emph{clustering
dynamic graphs}, i.e. graphs where the clustering corresponds to the partition
of nodes into natural groups based on the concept of density of edges within and between the clusters. The generator generates a time series
of random graphs $G_0, G_1, ..., G_n$, where $G_t$ emerges from $G_{t-1}$ via
successive atomic updates like for instance, the insertion of a vertex or the removal of an
edge. The generator dynamically monitors the ground truth clustering, and the probability
of the updates is chosen in such a way that the ground truth is
maintained while the randomness of the generated graph is kept.

Another recent proposal of a generator~\cite{mlg2018_42} of temporal graphs
results from an observation that small subgraph patterns in networks, called
\emph{network motifs} or \emph{graphlets}, are crucial indicators of the
structure and the evolution of the
graphs~\cite{Paranjape:2017:MTN:3018661.3018731}. For a given graph and a
predefined ordered list of structural atomic motifs the generator first computes
the distribution of the motifs in the graph. The distribution is then used to
generate a synthetic graph with the same features.


\subsection{Multi-Model Data}
With the dawn of Big Data and especially its Variety, another key 3V characteristic, new types of
database management systems have emerged. One of the most interesting ones
are the so-called \emph{multi-model databases}~\cite{Lu:2019:MDN:3341324.3323214} that enable to store and
thus query across structurally different data, including unstructured, semi-structured, and structured. There exist various types of
multi-model systems combining  distinct subsets of Big Data structures including graph data.
For example, OrientDB\footnote{\url{http://orientdb.com/orientdb/}} which has been mainly designed as an object DBMS currently supports graph, document,
key/value, and object models. Such type of DBMSs also needs a specific
data generator that would enable
to test new features and analyze efficiency of operations. However, since the
multi-model systems are in the context of Big Data rather new, there exist only
a few benchmarks targeting multi-model DBMSs (such as
Bigframe~\cite{journals/pvldb/KunjirKB14} or UniBench~\cite{conf/cidr/lu17})
with limited capabilities.

Another interesting approach to multi-model data is to adopt a unifying
expressive graph data model, so-called \emph{property graph data
model}~\cite{BFVY18}. Such a  model allows to specify multi-edges and list of
properties for the nodes. Synthetic graph generators for property graphs and its
companion standard graph query language~\cite{Angles18,AnglesABBFGLPPS18} are also needed in order to boost
their availability and adoption for different communities.

\subsection{Machine Learning Based Graph Generation}

With the advent of neural networks and specially generative adversarial networks
(GANs)~\cite{goodfellow2014generative}, several researchers have started to
explore their application to generate graphs. This is the case of
~\cite{simonovsky2018graphvae,kipf2016variational,grover2018graphite,li2018learning,you2018graphrnn},
which present several generative models to generate realistic graphs.  Such
techniques still suffer from several problems. For instance, some of them are
limited to learn from a single
graph~\cite{kipf2016variational,grover2018graphite} or generate small
graphs~\cite{simonovsky2018graphvae,you2018graphrnn,li2018learning}. The
technique proposed in~\cite{you2018graphrnn} is capable of generating graphs
with complex edge dependencies (e.g. community structure) and is not restricted
to graphs of a fixed size. However, there are still in general several open
challenges, including the capability of learning from and generating large graphs comparable
in size to those typically used for benchmarking, and robust generation
techniques with structural guarantees (e.g. degree distribution, clustering
coefficient, etc.).

\subsection{Privacy-Preserving Graph Generation}

A lot of work has been conducted on techniques for publishing social network
graphs with privacy guarantees~\cite{wu2010survey}. However, the topic of
generating social graphs with a realistic structure yet private has been barely explored.

Most of the existing work falls within the topic of graph generation with ``differential
privacy''~\cite{dwork2009differential} guarantees. More specifically, in
~\cite{wang2013preserving} the authors develop a differential privacy graph
generation approach based on the dK-graph generation
model~\cite{mahadevan2006systematic} that outperforms the Stochastic Kronecker
Graph Model~\cite{Leskovec:2005:RMT:2101235.2101254} in terms of the produced structural properties,
even though the results show that there is still room for improvement.

Following this line of research, recent work~\cite{qin2017generating} extends the notion of differential privacy
and propose an ``edge local differential privacy'' based graph generation
method. The proposed method allows generating privacy preserving synthetic social
graphs without the need of a centralized data curator, while preserving structural
properties more accurately than straw-hat methods such as Randomized
Neighbor Lists (based on randomized response~\cite{dwork2014algorithmic}) and
Degree-based Graph Generation (which perturbates the original graph degrees
using the Laplace mechanism~\cite{dwork2009differential}). Again, even though the
proposed technique outperforms the baselines, the results show that there is
still room for improving the structural properties of the
generated graph.

\section{Conclusion}
\label{sec:conclusion}

Graph data occur in a vast amount of distinct applications, such as biology, chemistry, physics, computer science, or social sciences, to name just a few. Graphs form one of the most complex data structures requiring specific and usually sophisticated approaches for processing and analysis. The history of graph theory,  that started from when these structures and their respective algorithms were studied, can be traced back to the 18th century.

With the recent dawn of Big Data there have  been more occurrences of large scale graphs where the efficiency   of  processing methods is critical.  Approaches that work for smaller scale graphs often cannot be used, the data need to be processed in a distributed way and hence the efficiency is influenced by other aspects, such as limits of data transport. In addition, distribution of graphs, especially for highly connected cases, is a difficult task. Thus extensive testing of these methods for graphs of various sizes and structural complexity is extremely important.

The aim of this survey was to provide a thorough overview and comparison of graph data generators. We do not limit ourselves to a single application domain, but we cover the currently most popular areas of graph data processing. We believe that this wide scope provides a uniquely useful insight into state-of-the-art tools as well as open issues for both researchers and practitioners.

\begin{acks}
The authors would like to thank Dr. Kamesh Madduri for consultations and suggestions on the covered areas.
The work of Sherif Sakr is funded by the European Regional Development Funds via the Mobilitas Plus programme (grant MOBTT75).
The work of Irena Holubov\'{a} was partially funded by the GA\v{C}R project no. 19-01641S.
\end{acks}

\bibliographystyle{ACM-Reference-Format-Journals}
\bibliography{biblio}

\received{June 2019}{June 2019}{June 2019}

\end{document}